\documentclass[preprint,12pt]{elsarticle} 

\usepackage[ruled]{algorithm2e}

\SetAlFnt{\small}
\SetAlCapFnt{\small}
\SetAlCapNameFnt{\small}
\SetAlCapHSkip{0pt}
\IncMargin{-\parindent}

\usepackage{epstopdf}
\usepackage{caption}
\usepackage{hyperref}

\journal{Information Systems}

\begin{document}

\begin{frontmatter}


\title{Aggregated 2D Range Queries on Clustered Points\tnoteref{t1,t2}}
\tnotetext[t1]{Funded in part by European Union's Horizon 2020 research and innovation programme under the Marie Sk{\l}odowska-Curie grant agreement No 690941, by Millennium Nucleus Information and Coordination in Networks ICM/FIC P10-024F (Chile), by MINECO (PGE and FEDER) Projects TIN2013-46238-C4-3-R and TIN2013-46801-C4-3-R (Spain), and also by Xunta de Galicia (GRC2013/053) (Spain). A preliminary partial version of this article appeared in
{\em Proc. SPIRE 2014}, pp.\ 215--226.}
\tnotetext[t2]{This is an Author's Original Manuscript of an article whose final and definitive form, the Version of Record, has been published in Information Systems [copyright Elsevier], available online at: http://dx.doi.org/10.1016/j.is.2016.03.004.}


\author[udc]{Nieves R. Brisaboa}
\author[enx]{Guillermo De Bernardo}
\author[uchile]{Roberto Konow}
\author[uchile]{Gonzalo Navarro}
\author[udec]{Diego Seco\tnoteref{t2}}
\tnotetext[t2]{Corresponding author: dseco@udec.cl. Tel.: +56 41 2204692; fax: +56 41 2221770}

\address[udc]{University of A Coru\~na, Campus de Elvi\~na, A Coru\~na, Spain}
\address[enx]{Enxenio S.L., Ba\~nos de Arteixo, A Coru\~na, Spain}
\address[uchile]{DCC, University of Chile, Beauchef 851, Santiago, Chile}
\address[udec]{University of Concepci\'on, Edmundo Larenas 219, Concepci\'on, Chile}

\begin{abstract}
Efficient processing of aggregated range queries on two-dimensional grids is a common requirement in information retrieval and data mining systems, for example in Geographic Information Systems and OLAP cubes. We introduce a technique to represent grids supporting aggregated range queries that requires little space when the data points in the grid are clustered, which is common in practice. We show how this general technique can be used to support two important types of aggregated queries, which are ranked range queries and counting range queries. Our experimental evaluation shows that this technique can speed up aggregated queries up to more than an order of magnitude, with a small space overhead.
\end{abstract}




\end{frontmatter}





\section{Introduction}


Many problems in different domains can be interpreted geometrically by modeling data records as multidimensional points and transforming queries about the original records into queries on the point sets \cite[Chapter 5]{Berg}. In 2D, for example, orthogonal range queries on a grid can be used to solve queries of the form \emph{``report all employees born between $y_0$ and $y_1$ who earn between $s_1$ and $s_2$ dollars''}, which are very common in databases. In the same way, other aggregated range queries (e.g., top-$k$, counting, quantile, majority, etc.) have proved to be useful for data analysis in various domains, such as Geographic Information Systems (GIS), OLAP databases, Information Retrieval, and Data Mining, among others \cite{NNRtcs13}. In GIS, aggregated range queries can facilitate decision making \cite{Harvey09} by counting, for example, the number of locations within a specific area for which the values of pollution are above a threshold. Similarly, top-$k$ range queries on an OLAP database\footnote{The support of more than two dimensions is essential in OLAP databases. We discuss the extension to multi-dimensional structures in the conclusions.} of sales can be used to find the sellers with most sales in a time slice. In this example, the two dimensions of the grid are seller ids (arranged hierarchically in order to allow queries for sellers, stores, areas, etc.) and time (in periods of hours, days, weeks, etc.), and the weights associated to the data points are the amount of sales made by a seller during a time slice. Thus, the query asks for the $k$ heaviest points in some range $Q=[i_1,i_2]\times[t_1,t_2]$ of the grid.

The approach of modeling problems using a geometric formulation is well-known. There are many classical representations that support the queries required by the model and solve them efficiently. Range trees \cite{Bentley79} and $kd$-trees \cite{Ben75} are two paradigmatic examples. Some of these classical data structures are even optimal both in query time and space. However, such classical representations usually do not take advantage of the distribution of the data in order to reduce the space requirements. When dealing with massive data, which is the case of some of the aforementioned data mining applications, the use of space-efficient data structures can make the difference between maintaining the data in main memory or having to resort to (orders of magnitude slower) external memory. 

In this work we consider the case where we have clustered points in a 2D grid, which is a common scenario in domains such as Geographic Information Systems, Web graphs, social networks, etc. There are some well-known principles that hold in most scenarios of that kind. Two examples are Tobler's first law of geography \cite{tobler}, which states that near things are more related than distant things, and the locality of reference for time-dependent data. This is also the case in Web graphs \cite{BoVWFI}, where clusters appear when the Web pages are sorted by URL. We take advantage of these clusters in order to reduce the space of the data structures for aggregated 2D range queries.

The $K^2$-tree \cite{ktree} (a space-efficient version of the classical Quadtree) is a good data structure to solve range queries on clustered points and it has been extensively evaluated in different domains \cite{Alvarez-GarciaB15,dBABNPspire13.3,CaroRB15}. We introduce a general technique to extend this data structure in order to support aggregated range queries. We then illustrate its potential by instantiating the technique in two emblematic cases: range counting and ranked (MAX/MIN) queries within a 2D range. 

The paper is organized as follows. First, we introduce basic concepts and related work in Sections \ref{sec:basics} and \ref{sec:relwork}, respectively. In Section \ref{sec:generaltech} we describe the general technique to extend the $K^2$-tree to solve different aggregated range queries on grids. Two paradigmatic examples of such queries are described in Section \ref{sec:topk} (ranked range queries) and Section \ref{sec:rangecounting} (range counting queries). Section \ref{sec:experiments} presents an exhaustive empirical evaluation of the proposed solutions. Finally, Section \ref{sec:conclusions} concludes and sketches some interesting lines of future work.

\section{Basic Concepts}\label{sec:basics}
\subsection{Aggregated queries on clustered points}

\begin{figure}[t]
		\centering
		\begin{minipage}[t]{0.31\textwidth}
			\centering
			 \includegraphics[]{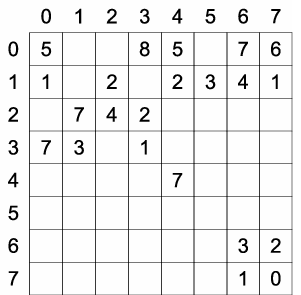}
			 \captionof{figure}{Weighted matrix.}\label{fig:matrix}
		\end{minipage}
	  	\begin{minipage}[t]{0.31\textwidth}
			\centering
			 \includegraphics[]{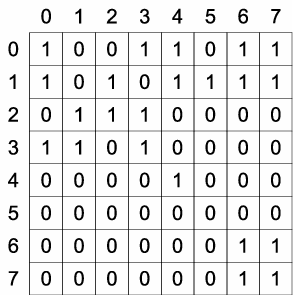}
			 \captionof{figure}{Binary matrix.}\label{fig:matrix_bin}
		\end{minipage}
	  	\begin{minipage}[t]{0.31\textwidth}
			\centering
			 \includegraphics[]{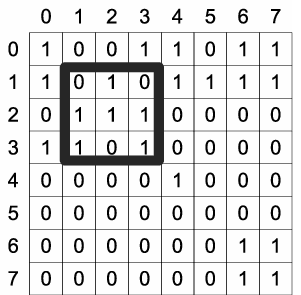}
			 \captionof{figure}{Range query.}\label{fig:matrix_qry}
		\end{minipage}
\end{figure}

We consider two dimensional grids with $n$ columns and $m$ rows, where each cell $a_{ij}$ can either be empty or contain a weight in the range $[0,d-1]$ (see Fig. \ref{fig:matrix}). For some problems, we will omit the weights and just consider the non-empty cells, which can be represented as a binary matrix (Fig. \ref{fig:matrix_bin}) in which each cell contains a 1 (if there is a weighted point in the original matrix) or a 0 (in other case). 

Let $t$ be the number of 1s in the binary matrix (i.e., the number of weighted points). If we can partition the $t$ points into $c$ clusters, not necessarily disjoint and $c << t$, we will say that the points are clustered. This definition is used by Gagie et al.\ \cite{GHKNPSdcc15.2} to show that in such case a Quadtree needs only $O(c\log u + \sum_i t_i \log l_i)$ bits, where $u=\max(n,m)$, and $t_i$ and $l_i$ are the number of points and the diameter of cluster $i$, respectively.

A range query $Q=[x_1,x_2]\times[y_1,y_2]$ defines a rectangle with all the columns in the range $[x_1,x_2]$ and the rows in $[y_1,y_2]$ (see Fig. \ref{fig:matrix_qry}). An aggregated range query defines, in addition to the range, an aggregate function that must be applied to the data points in the query range. Examples of aggregated queries are $\mathcal{COUNT}(Q)$, which counts the number of data points in the query range, $\mathcal{MAX/MIN}(Q)$, which computes the maximum (alt. minimum) value in the query range, and its generalization top-$k$, which retrieves the $k$ lightest (alt. heaviest) points in the query range. These top-$k$ queries are also referred to in the literature as ranked range queries. For the range query $q$ in Fig. \ref{fig:matrix_qry} the result of $\mathcal{COUNT}(q)$ is 6, $\mathcal{MAX}(q)$ returns 7, $\mathcal{MIN}(q)$ returns 1, and the top-$3$ heaviest elements are 7, 4, and 3.

There are other interesting data-analysis queries on two-dimensional grids. For example, $\mathcal{QUANTILE}(Q,a)$ returns the $a$-th smallest value in $Q$, and $\mathcal{MAJORITY}(Q,\alpha)$ retrieves those values in $Q$ that appear with relative frequency larger than $\alpha$. These and other queries have been studied by Navarro et al.\ \cite{NNR13}, who introduce space-efficient data structures with good time performance. We restrict ourselves to an emblematic subset of these queries, and propose data structures that are even more space-efficient when the points in the set are clustered. 

\subsection{Rank and select on bitmaps}
Two basic primitives used by most space-efficient data structures are rank and select on bitmaps. Let $B[1,n]$ be a sequence of bits, or a bitmap. We define operation $rank_b(B,i)$ as the number of occurrences of $b \in \{0,1\}$ in $B[1,i]$, and $select_b(B,j)$ as the position in $B$ of the $j$-th occurrence of $b$. $B$ can
be represented using $n+o(n)$ bits \cite{Jac89,Cla96}, so that both operations are
solved in constant time.  These operations have proved very efficient in practice \cite{CN08}. In addition, when the bitmaps are compressible, it is possible to reduce the space and still support these operations in constant time \cite{RRR02}.

\subsection{Wavelet tree and discrete grids}\label{subsec:wt}


An elegant generalization of rank and select queries to an arbitrary alphabet $\Sigma$ of size $\sigma$ is provided by the wavelet tree \cite{GGV03}. Given a sequence $S$ over the alphabet $\Sigma$, the wavelet tree supports rank, select and access in $O(\log \sigma)$ time with $n\log \sigma+o(n\log \sigma)$ bits. The wavelet tree is a complete binary tree, in which each node represents a range $R\subseteq [1,\sigma]$ of the alphabet $\Sigma$, its left child represents a subset $R_\ell \subset R$ and the right child the subset $R_r = R\setminus R_\ell$. Every node representing subset $R$ is associated with a subsequence $S'$ of the input sequence $S$ composed of the elements whose values are in $R$. The node only stores a bitmap of length $|S'|$ such that a $0$ bit at position $i$ means that $S'[i]$ belongs to $R_\ell$, and a $1$ bit means that it belongs to $R_r$. The three basic operations require to traverse the tree from the root to a leaf (for rank and access) or from a leaf to the root (for select) via rank and select operations on the bitmaps stored at the nodes of the wavelet tree (those bitmaps are then represented with the techniques cited above).

The wavelet tree can also be used to represent grids \cite{Cha88,MN06}. An $n \times m$ grid with $n$ points, exactly one per column (i.e., $x$ values are unique), can be represented using a {\em wavelet tree}. In this case, this is a perfect balanced binary tree of height $\lceil \log m\rceil$ where each node corresponds to a contiguous range of values $y \in [1,m]$ and represents the points falling in that $y$-range, sorted by increasing $x$-coordinate. The root represents $[1,m]$ and the two children of each node split its $y$-range by half. The leaves represent a single $y$-coordinate. Each internal node stores a bitmap, which tells whether each point corresponds to its left or right child. Using $rank$ and $select$ queries on the bitmaps, the wavelet tree uses $n\log m + o(n\log m)$ bits, and can count the number of points in a range in $O(\log m)$ time, because the query is decomposed into bitmap ranges on at most 2 nodes per wavelet tree level (see Section \ref{sec:relwork}). Any point can be tracked up (to find its $x$-coordinate) or down (to find its $y$-coordinate) in $O(\log m)$ time as well.

When the grids may contain more than one point per column, an additional bitmap $B$ is used to map from the original domain to a new domain that has one point per column. This bitmap stores, for each column, a bit 1 followed by as many zeros as the number of points in such column. Then, a range of columns $[c_s,c_e]$ in the original domain can be mapped to a new range $[select_1(B,c_s)-c_s+1,select_1(B,c_e+1)-c_e-1]$. If the grid is very sparse and/or the distribution of the data points is very skewed, this bitmap can be compressed with RRR \cite{RRR02} or the sd-bitvector \cite{OkanoharaS07}.

\subsection{$K^2$-trees}\label{sec_k2tree}
\begin{figure}[t]
 \centering
 \includegraphics[width=\textwidth]{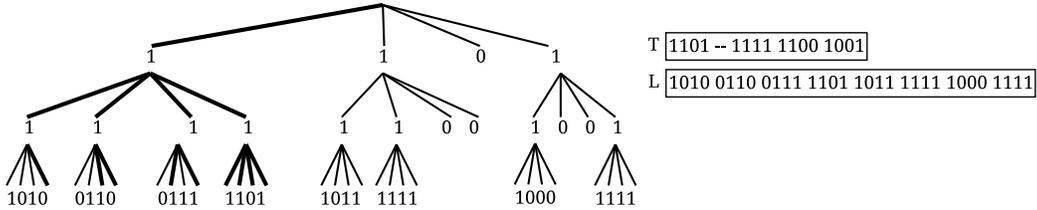}
\caption{On the left, the conceptual $K^2$-tree for the binary matrix in Fig. \ref{fig:matrix_bin} (highlighted edges are traversed when computing the range query in Fig. \ref{fig:matrix_qry}). On the right, the two bitmaps that are actually used to store the tree.}
\label{fig:k2tree}
\end{figure}

The $K^2$-tree~\cite{ktree} is a data structure designed to compactly represent sparse binary matrices (which can also be regarded as point grids).
The $K^2$-tree subdivides the matrix into $K^2$ submatrices of equal size. In this regard, when $K=2$, the $K^2$-tree performs the same space partitioning of the traditional Quadtree. The submatrices are considered left-to-right and top-to-bottom (i.e., in Morton order), and each is represented with a bit, set to 1 if the submatrix contains at least one non-zero cell. Each node whose bit is 1 is recursively decomposed, subdividing its submatrix into $K^2$ children, and so on. The subdivision ends when a fully-zero submatrix is found or when we reach the individual cells. A $K^2$-tree for the running example is shown in Fig. \ref{fig:k2tree}.

The $K^2$-tree is stored in two bitmaps: $T$ stores the bits of all the levels 
except the last one, in a level-order traversal, and $L$ stores the bits of 
the last level (corresponding to individual cells). Given a node whose bit is
at position $p$ in $T$, its children nodes are located after position $rank_1(T,p) \cdot K^2$. This property enables $K^2$-tree traversals using just $T$ and $L$. 

The worst-case space, if $t$ points are in an $n \times n$ matrix, is $K^2\,t \log_{K^2} \frac{n^2}{t} (1+o(1))$ bits. This can be reduced to $t \log \frac{n^2}{t}  (1+o(1))$ if the bitmaps are compressed. This is similar to the space achieved by a wavelet tree, but in practice $K^2$-trees use much less space when the points are clustered. Gagie et al.\ \cite{GHKNPSdcc15.2} show that this quadtree-like partitioning results in $O(c\log n+\sum_i t_i \log l_i)$ bits, when the $t$ points can be partitioned into $c$ clusters with $t_i, \ldots, t_c$ points and diameters $l_1,\ldots,l_c$. Therefore, the $K^2$-tree is competitive in domains where such clusters arise, for example in Web graphs or social networks.

Among other types of queries (such as direct/reverse neighbors, check edge, etc.), the $K^2$-tree can answer range queries with multi-branch top-down traversal of the tree, following only the branches that overlap the query range. This is illustrated in Algorithm \ref{alg:range} (adapted from \cite[Alg.\ 1]{ktree}), which solves the query $Q= [x_1,x_2] \times [y_1,y_2]$ by invoking $Range(n,x_1,x_2,y_1,y_2,0,0,-1)$.
 To show an example, in Fig. \ref{fig:k2tree} the edges traversed in the computation of the range query in Fig. \ref{fig:matrix_qry} are highlighted. As mentioned above, this traversal is computed via rank queries on $T$. While this algorithm has no good worst-case time guarantees, in practice times are competitive.

\begin{algorithm}[t]
    \caption{{\bf Range}$(n,x_1,x_2,y_1,y_2,d_p,d_q,p)$ lists all non-empty cells in {$[x_1,x_2]$} $\times$ {$[y_1,y_2]$ with a $k^2$-tree}}
    \eIf (\tcc*[h]{leaf}){$p \ge |T|$}{
    \lIf{$L[p-|T|]=1$}{output $(d_p,d_q)$}
    }
    (\tcc*[h]{internal node})
    {
    \If{$p = -1 ~\mathbf{or}~ T[p]=1$}{
    $y \leftarrow rank_1(T,p) \cdot k^2$ \\
    \For{$i\leftarrow\lfloor x_1 / (n/k) \rfloor \ldots \lfloor x_2 / (n/k) \rfloor$}{
        \lIf{$i\leftarrow\lfloor x_1 / (n/k) \rfloor$}{
		{$x_1' \leftarrow x_1~\textrm{mod}~(n/k)$}
        }
        \lElse{$x_1' \leftarrow 0$}
        \lIf{$i\leftarrow\lfloor x_2 / (n/k) \rfloor$}{
        {$x_2' \leftarrow x_2~\textrm{mod}~(n/k)$}
        }
        \lElse{$x_2' \leftarrow (n/k)-1$}
        \For{$j\leftarrow\lfloor y_1 / (n/k) \rfloor \ldots \lfloor y_2 / (n/k) \rfloor$}{
        \lIf{$j=\lfloor y_1 / (n/k) \rfloor$}{
		{$y_1' \leftarrow y_1~\textrm{mod}~(n/k)$}
        }
        \lElse{$y_1' \leftarrow 0$}
        \lIf{$j=\lfloor y_2 / (n/k) \rfloor$}{
        {$y_2' \leftarrow y_2~\textrm{mod}~(n/k)$}
        }
        \lElse{$y_2' \leftarrow (n/k)-1$}
        {{\bf Range}$(n/k,x_1',x_2',y_1',y_2',d_p+(n/k)\cdot i, d_q+(n/k)\cdot j,y+k\cdot i + j)$}
    }
    }
    }
    }
    \label{alg:range}
\end{algorithm}


\subsection{Treaps, priority search trees and ranked range queries}

A \emph{treap} \cite{SA96} is a binary search tree whose $n$ nodes have two
attributes: \emph{key} and \emph{priority}. The treap maintains the binary
search tree invariants for the keys and the heap invariants for the
priorities, that is, the key of a node is larger than those in its left
subtree and smaller than those in its right subtree, whereas its priority 
is not smaller than those in its subtree. The treap does not guarantee
logarithmic height, except on expectation if priorities are independent of
keys \cite{MS97}. A treap can also be regarded as the Cartesian tree \cite{Vui80} of the sequence of priorities once the values are sorted by keys. The succinct
representations of the Cartesian tree topology \cite{FH11} are called range
maximum query (RMQ) data structures, use just $2n+o(n)$
bits, and are sufficient to find the maximum in any range of the sequence. By
also storing the priority data, they can answer top-$k$ queries in $O(k \log k)$ or $O(k \log\log n)$ time. The treap can also be used to compress the
representation of keys and priorities \cite{KNCLOsigir13}.
Similar data structures for two or more dimensions are convenient only for dense
grids (full of points) \cite{GIKRR11}.

The \emph{priority search tree} \cite{mc85} is somewhat similar, but it is balanced. In this case, a node is not the one with highest priority in its subtree, but the highest-priority element is stored in addition to the element at the node, and removed from the subtree. Priority search trees can be used to solve 3-sided range queries on $t$-point grids, returning $k$ points in time $O(k+\log t)$. This has been used to add rank query capabilities to several index data structures such as suffix trees and range trees \cite{iwona05}.


\section{Related Work}\label{sec:relwork}

Navarro et al.\ \cite{NNR13} introduce compact data structures for various
queries on two-dimensional weighted points, including range top-$k$ queries and range counting queries.
Their solutions are based on wavelet trees. For range top-$k$ queries, the bitmap of each node of the wavelet tree is enhanced as follows: Let $x_1,\ldots,x_r$ be
the points represented at a node, and $w(x)$ be the weight of point $x$. Then, 
a RMQ data structure built on $w(x_1),\ldots,w(x_r)$
is stored together with the bitmap. Such a structure uses $2r+o(r)$ bits and
finds the maximum weight in any range $[w(x_i),\ldots,w(x_j)]$ in constant
time \cite{FH11} and without accessing the weights themselves. Therefore, the
total space becomes $3n\log m + o(n\log m)$ bits.

To solve top-$k$ queries on a grid range $Q= [x_1,x_2] \times [y_1,y_2]$, we 
first traverse the wavelet tree to identify the $O(\log m)$ bitmap intervals 
where the points in $Q$ lie. The heaviest point in $Q$ in each bitmap
interval is obtained with an RMQ, but we need to obtain the actual priorities
in order to find the heaviest among the $O(\log m)$ candidates. The priorities
are stored sorted by $x$- or $y$-coordinate, so we obtain each one in $O(\log
m)$ time by tracking the point with maximum weight in each interval. Thus a
top-1 query is solved in $O(\log^2 m)$ time. For a top-$k$ query, we must
maintain a priority queue of the candidate intervals and, each time the next
heaviest element is found, we remove it from its interval and reinsert in
the queue the two resulting subintervals. The total query time is
$O((k+\log m)\log (km))$.
It is possible to reduce the time to $O((k+\log m)\log^\epsilon m)$ time and
$O(\frac{1}{\epsilon}n\log m)$ bits, for any constant $\epsilon>0$
\cite{NNsoda12}, but the space usage is much higher, even if linear.

Wavelet trees can also compute range counting queries in $O(\log m)$ time with $n\log m+o(n \log m)$ bits. The algorithm to solve range counting queries on a grid range $Q= [x_1,x_2] \times [y_1,y_2]$ also starts by traversing the wavelet tree to identify the $O(\log m)$ bitmap intervals where the points in $Q$ lie, but then it just adds up all the bitmap interval lengths.

A better result, using multi-ary wavelet trees, was introduced by Bose et al.\ \cite{BoseHMM09}. They match the optimal $O(\log n / \log\log n)$ time using just $n\log n +o(n\log n)$ bits on an $n \times n$ grid. Barbay et al.\ \cite{BCNic13} extended the results to $n \times m$ grids. This query time is optimal within space $O(n~polylog(n))$ \cite{Patrascu07}.

\section{Augmenting the $K^2$-tree}\label{sec:generaltech}
In this section we describe a general technique that can be used to solve aggregated range queries on clustered points. We then present two applications of this technique to answer two paradigmatic examples, ranked range queries and range counting queries. These examples illustrate how to adjust and tune the general technique for particular operations.

Let $M[n\times n]$ be a matrix in which cells can be empty or contain a weight in the range $[0, d-1]$ and let $BM[n\times n]$ be a binary matrix in which each cell contains a zero or a one. Matrix $BM$ represents the topology of $M$, that
is, $BM[i][j]=1$ iff $M[i][j]$ is not empty.

We store separately the topology of the matrix and the weights associated with the non-empty cells. For the topology, we use a $K^2$-tree representation of $BM$ (recall Section \ref{sec_k2tree}), which will take advantage of clustering.

A level-wise traversal of the $K^2$-tree can be used to map each node to a position in an array of aggregated values, which stores a {\em summary} of the weights in the submatrix of the node. Thus the position where the aggregated value of a node is stored is easily computed from the node position in $T$. 

The specific value of this summary depends on the operation. For example, for ranked range queries (Section \ref{sec:topk}) the summary represents the maximum weight in the corresponding submatrix, whereas for counting queries (Section \ref{sec:rangecounting}), it represents the number of non-empty cells in the submatrix. However, a common property is that the value associated with a node aggregates information about its $K^2$ children. Therefore, we use a sort of differential encoding \cite{CNS14} to encode the values of each node with respect to the value of its parent. In other words, the information of a node (such as its summary and number of children) is used to represent the information of its children in a more compact way. In order to access the original (non-compressed) information of a node we need to first access its parent (i.e., the operations in this technique are restricted to root-to-leaf traversals).

To summarize, we use a $K^2$-tree to represent the topology of the data and augment each node of such a tree with additional values that represent the aggregated information related with the operation to be supported. These aggregated values are differentially encoded with respect to information of the parent node in order to store them in reduced space. In the following sections we show how both the $K^2$-tree and the differentially encoded values can be tuned to efficiently solve two types of queries.

Finally, note that we present our results for matrices of size $n \times n$. This does not lose generality, as we can extend a matrix $M'[n \times m]$ with zeros to complete a square matrix $M[n \times n]$ (w.l.o.g. we assume $m \leq n$). As the topology of the matrix is represented with a $K^2$-tree, this does not cause a significant overhead because the $K^2$-tree is efficient to handle large areas of zeros. Actually, we round $n$ up to the next power of $K$ \cite{ktree}.

\section{Answering Ranked (Max/Min) Range Queries}\label{sec:topk}

We present a first application of the general technique described in the previous section, to solve ranked range queries. We present the case of $\mathcal{MAX}$ queries, but the results are analogous for $\mathcal{MIN}$ queries. We name this data-structure $K^2$-treap, as it conceptually combines a $K^2$-tree with a treap data structure. 

Consider a matrix $M[n \times n]$ where each cell can either be empty or contain a weight in the range $[0,d-1]$. We consider a quadtree-like recursive partition of $M$ into $K^2$ submatrices, the same performed in
the $K^2$-tree with binary matrices. We build a conceptual $K^2$-ary tree
similar to the $K^2$-tree, as follows: the root of
the tree will store the coordinates of the cell with the maximum weight of the
matrix, and the corresponding weight. Then the cell just added to the tree is
marked as \emph{empty},
deleting it from the matrix. If many cells share the maximum weight,
we pick any of them. Then, the matrix is conceptually decomposed into $K^2$
equal-sized submatrices, and we add $K^2$ children nodes to the root of the
tree, each representing one of the submatrices. We repeat the assignment
process recursively for each child, assigning to each of them the coordinates
and value of the heaviest cell in the corresponding submatrix and removing the
chosen point. The procedure continues recursively on each branch until we
reach the cells of the matrix, or we find a completely empty submatrix 
(either because the submatrix was initially
empty or because we emptied it by successively extracting heaviest points).

\begin{figure}[]
 \centering
 \includegraphics[width=\textwidth]{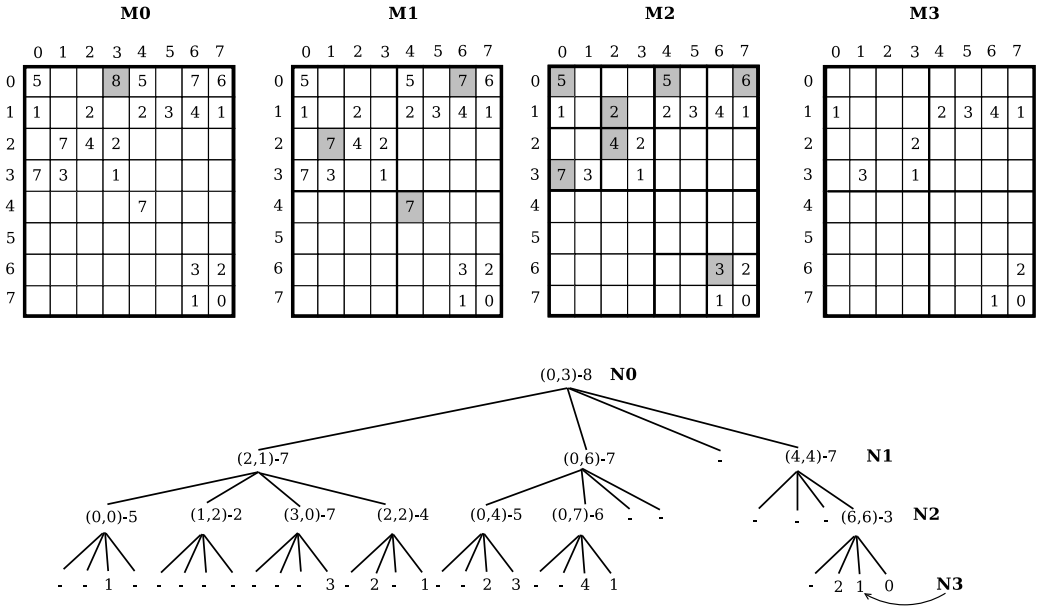}
\caption{Example of a $K^2$-treap construction for the matrix in Fig. \ref{fig:matrix}. At the top, $M_i$ represents the state of the matrix at level $i$ of the decomposition. On the bottom, the conceptual $K^2$-treap.}
\label{fig:ck2treap}
\end{figure}

Fig.~\ref{fig:ck2treap} shows an example of $K^2$-treap construction, for
$K=2$. On the top of the image we show the state of the matrix at each level of
the decomposition. $M0$ represents the original matrix, where the maximum value is
highlighted. The coordinates and value of this cell are stored in the root of
the tree. In the next level of the decomposition (matrix $M1$) we find the maximum
values in each quadrant (notice that the cell assigned to the root has already
been removed from the matrix) and assign them to the children of the root
node. The process continues recursively, subdividing each matrix into $K^2$
submatrices. The cells chosen as local maxima are highlighted in the matrices
corresponding to each level, except in the last level where
all the cells are local maxima. Empty submatrices are marked in the tree with the symbol ``\texttt{-}''.

\subsection{Local maximum coordinates}

The data structure is represented in three parts: The coordinates of the local maxima,
the weights of the local maxima, and the tree topology.

\begin{figure}[t]
 \centering
 \includegraphics[width=\textwidth]{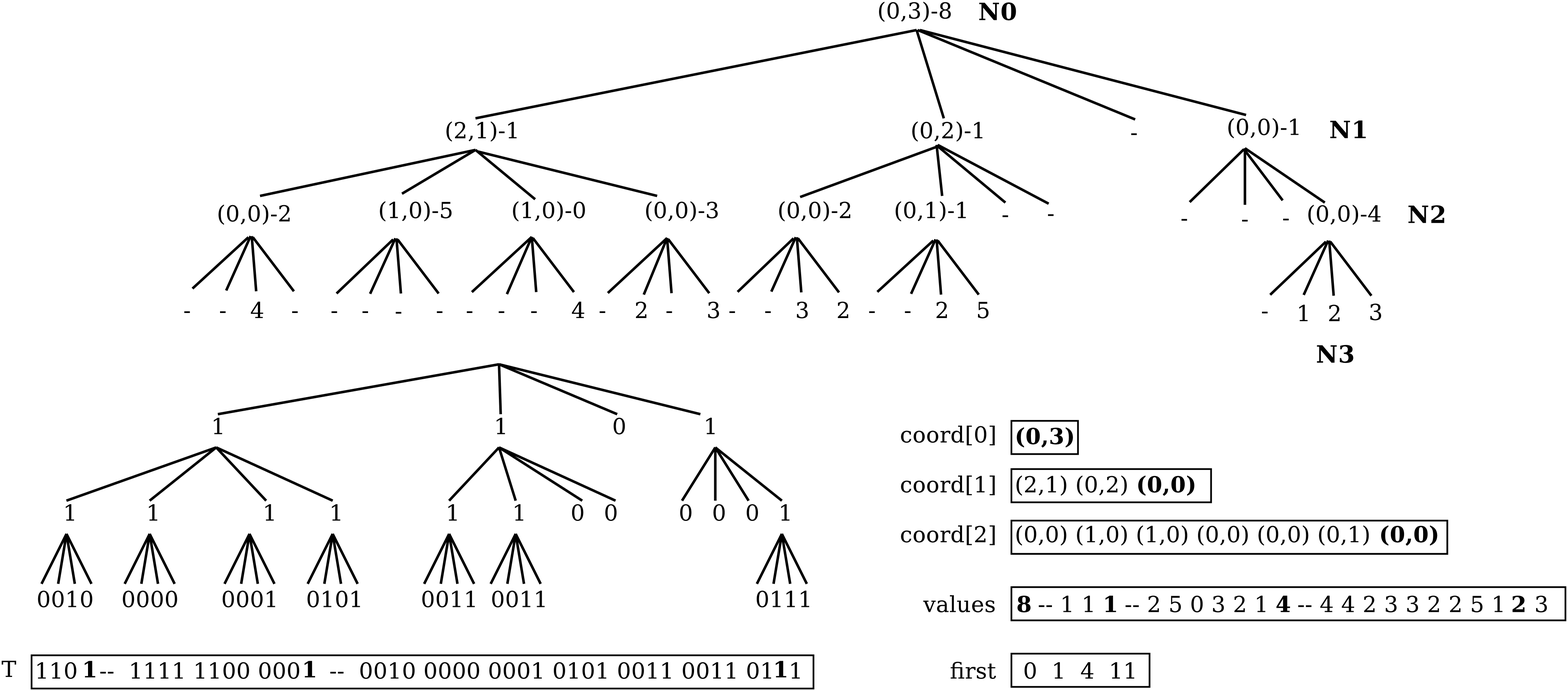}
\caption{Storage of the conceptual tree in our data structures. On the top, the differentially encoded conceptual $K^2$-treap. On the bottom left, the conceptual $K^2$-tree that stores the topology of the matrix, and its bitmap implementation $T$. On the bottom right, the local maximum values.}
\label{k2treapReal}
\end{figure}

The conceptual $K^2$-treap is traversed
level-wise, reading the sequence of cell coordinates from left to right in
each level. The sequence of coordinates at each level $\ell$ is stored in a
different sequence $coord[\ell]$. The coordinates at each level $\ell$ of the
tree are transformed into an offset in the corresponding submatrix,
representing each $c_i$ as $c_i~\mathrm{mod}~(n / K^\ell)$ using $\lceil
\log(n) - \ell\log K\rceil$ bits.
For example, in Fig.~\ref{k2treapReal} (top) the coordinates of node $N1$ have
been transformed from the global value $(4,4)$ to a local offset $(0,0)$. In
the bottom of Fig.~\ref{k2treapReal} we highlight the coordinates of nodes $N0$, $N1$ 
and $N2$ in the corresponding $coord$ arrays. In the last level all nodes represent single 
cells, so there is no $coord$ array in this level.
With this representation, the worst-case space for storing $t$ points 
is $\sum_{\ell=0}^{\log_{K^2}(t)} 2K^{2\ell}\log\frac{n}{K^\ell} =
t\log\frac{n^2}{t} (1+O(1/K^2))$, that is, the same as if we stored the points
using the $K^2$-tree.

\subsection{Local maximum values}
The maximum value in each node is
encoded differentially with respect to the maximum of its parent 
node~\cite{CNS14}. The result of the differential encoding is a new sequence
of non-negative values, smaller than the original. Now the $K^2$-treap is 
traversed level-wise and the complete sequence of values is stored in a single 
sequence named $values$. To exploit the small values while allowing efficient 
direct access to the array, we represent $values$ with Direct Access 
Codes (DACs)~\cite{BLN12}. Following the example in Fig.~\ref{k2treapReal}, the value
 of node $N1$ has been transformed from 7 to $8-7=1$. The bottom of the figure depicts the 
 complete sequence $values$. We also store a small array 
$\mathit{first}[0,\log_{K^2} n]$ that stores the offset in $values$ where each level starts.

\subsection{Tree structure}
We separate the structure of the tree from the values
stored in the nodes. The tree structure of the $K^2$-treap is stored in a
$K^2$-tree. Fig.~\ref{k2treapReal} shows the $K^2$-tree representation of
the example tree, where only cells with value are labeled with a 1. We will
consider a $K^2$-tree stored in a single bitmap $T$ with $rank$ support, that
contains the sequence of bits from all the levels of the tree. Our
representation differs from a classic $K^2$-tree (which uses two bitmaps $T$
and $L$ and only adds rank support to $T$) because we will need to perform
rank operations also in the last level of the tree. The other difference is
that points stored separately are removed from the grid. Thus, we save the
$K^2$-tree space needed to store those removed points. Our analysis above
shows that, in a worst-case scenario, the saved and the extra space cancel
out each other, thus
storing those explicit coordinates is free in the worst case.

\subsection{Query processing}
\subsubsection{Basic navigation}
To access a cell $C=(x,y)$ in the $K^2$-treap we start accessing the
$K^2$-tree root. The coordinates and weight of the element stored at the 
root node are $(x_0,y_0)=coord[0][0]$ and $w_0=values[0]$. If $(x_0,y_0)=C$, we
return $w_0$ immediately.
Otherwise, we find the quadrant where the cell
would be located and navigate to that node in the $K^2$-tree.
Let $p$ be the position
of the node in $T$. If $T[p]=0$ we know that the complete submatrix is empty and
return immediately. Otherwise, we need to find the coordinates and weight of
the new node. Since only nodes set to 1 in $T$ have coordinates and weights, we
compute $r=rank_1(T,p)$. The value of the current node will be at
$values[r]$, and its coordinates at $coord[\ell][r-\mathit{first}[\ell]]$, where $\ell$ 
is the current level. We rebuild the absolute value and coordinates, 
$w_1$ as $w_0 - values[r]$ and $(x_1,y_1)$ by adding the current submatrix offset
to $coord[\ell][r-\mathit{first}[\ell]]$. If $(x_1,y_1)=C$ we return $w_1$, otherwise 
we find again the appropriate quadrant in the current submatrix where $C$ 
would be located, and so on. 
The formula to find the children is identical to that of the 
$K^2$-tree. The process is repeated recursively until we find a 0 bit in the
target submatrix, we find a 1 in the last level of the $K^2$-tree, or we find 
the coordinates of the cell in an explicit point.

\subsubsection{Top-$k$ queries}

The process to answer top-$k$ queries starts at the root of the tree. Given a
range $Q= [x_1,x_2] \times [y_1, y_2]$, the process initializes an empty
max-priority queue and inserts the root of the $K^2$-tree. The priority queue 
stores, in general, $K^2$-tree nodes sorted by their associated maximum weight
(for the root node, this is $w_0$).
Now, we iteratively extract the first element from the priority queue (the first time 
this is the root). If the coordinates of its maximum element fall inside 
$Q$, we output it as the next answer. In either case,
we insert all the children of the extracted node whose submatrix intersects
with $Q$, and iterate. The process finishes when $k$ results have been found
or when the priority queue becomes empty (in which case there are less than $k$ 
elements in $Q$). 

\subsubsection{Other supported queries}
The $K^2$-treap can also answer basic range queries (i.e., report all the
points that fall in $Q$). This is similar to the procedure on a $K^2$-tree,
where the submatrices that intersect $Q$ are explored in a depth-first manner.
The only difference is that we must also check whether the explicit points
associated to the nodes fall within $Q$, and in that case report those as
well. Finally, we can also answer {\em interval queries}, which ask for all
the points in $Q$ whose weight is in a range $[w_1,w_2]$. To do this, we 
traverse the tree as in a top-$k$ range query, but we only output weights
whose value is in $[w_1,w_2]$. Moreover, we discard submatrices whose maximum
weight is below $w_2$.


\section{Answering Range Counting Queries}\label{sec:rangecounting}
Consider a binary matrix $BM[n \times n]$ where each cell can either be empty or contain data\footnote{It is easy to allow having more than one point per cell, by using the aggregated sums described in Section \ref{subsub_other}.}. In this case, a $K^2$-tree can be used to represent $BM$ succinctly while supporting range queries, as explained in Section \ref{sec_k2tree}. Obviously, the algorithm presented for range reporting can be optimized to count the number of elements in a range, instead of reporting such elements. In this section, we show how to augment the $K^2$-tree with additional data to further optimize those range counting queries. In Section \ref{sec:experiments} we show that this adds a small overhead in space, while drastically reducing the running time of those queries.

\subsection{Augmenting the $K^2$-tree}
\begin{figure}[t]
 \centering
 \includegraphics[width=\textwidth]{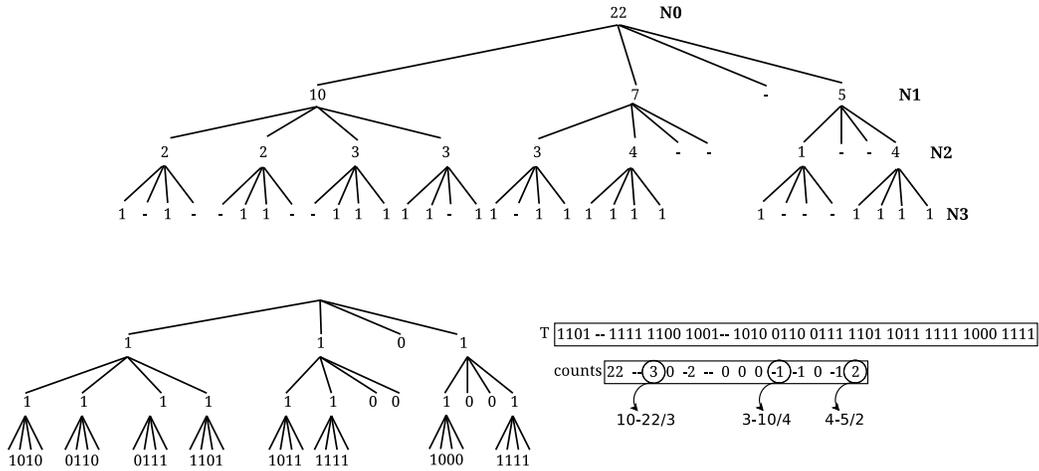}
\caption{Storage of the conceptual tree in our data structures. On the top, the conceptual $K^2$-treap for the binary matrix in Fig. \ref{fig:matrix_bin}. On the bottom left, the conceptual $K^2$-tree that stores the topology of the matrix. On the bottom right, the bitmap that implements the $k^2$-tree, $T$, and the sequence of differentially encoded counting values, $counts$.}
\label{fig:rck2tree}
\end{figure}

The augmented data structure stores additional information to speed up range counting queries. In Fig. \ref{fig:rck2tree} we show a conceptual $K^2$-tree in which each node has been annotated with the number of elements in its corresponding submatrix. Note that this is the same example of Fig.\ \ref{fig:ck2treap}, considering as non-empty cells those with weight larger than 0.

This conceptual tree is traversed level-wise, reading the sequence of counts from left to right at each level. All these counts are stored in a sequence $counts$ using a variant of the differential encoding technique presented in Section \ref{sec:generaltech}. Let $v$ be a node of the $K^2$-tree, $children(v)$ the number of children of $v$, and $count(v)$ the number of elements in the submatrix represented by $v$. Then, $\overline{count(v')}=\frac{count(v)}{children(v)}$ represents the expected number of elements in the submatrix associated with each child $v'$ of $v$, assuming a uniform distribution. Thus, the count of each node $v'$ is stored as the difference of the actual count and its expected value. In the running example, the root has three children and there are 22 elements in the matrix. Each of the corresponding submatrices is expected to contain $\lfloor 22/3 \rfloor = 7$ elements whereas they actually contain 10, 7 and 5 elements, respectively. Hence, the differential encoding stores $10-7=3$, $7-7=0$, and $5-7=-2$.

The result of this differential encoding is a new sequence of values smaller than the original, but which may contain negative values. In order to map this sequence, in a unique and reversible way, to a sequence of non-negative values we use the folklore \emph{overlap and interleave} scheme, which maps a negative number $-i$ to the $i^{th}$-odd number ($2i-1$) and a positive number $j$ to the $j^{th}$ even number ($2j$). Finally, to exploit the small values while allowing efficient direct access to the sequence, we represent $counts$ with DACs~\cite{BLN12}.

As $counts$ corresponds with a level-wise traversal of the $K^2$-tree, it is not necessary to store this additional information for all the levels of the tree. In this way, we provide a parametrized implementation that sets the space overhead by defining the number of levels for which counting information is stored. This provides a space-time trade-off we later study experimentally.

\subsubsection{Range counting queries}

The base of the range counting algorithm is the range reporting algorithm of the $K^2$-tree, Algorithm \ref{alg:range}. We modify this divide-and-conquer algorithm in order to navigate both the $K^2$-tree and $counts$ at the same time. Given a query range $Q=[x_1,x_2] \times [y_1,y_2]$, we start accessing the $K^2$-tree root, and set $c_0 = counts[0]$ and $result=0$. Then, the algorithm explores all the children of the root that intersect $Q$ and decodes the counting value of the node from the $counts$ sequence and the absolute counting value of the root. The process continues recursively for each children until it finds a completely empty submatrix, in which case we do not increment $result$, or a matrix completely contained inside $Q$, in which case we increment result with the counting value of such matrix.

To clarify the procedure, let us introduce some notation and basic functions. We name the nodes with the position of their first bit in the bitmap $T$ that represents the $K^2$-tree. Then, $v=0$ is the root, $v=K^2$ is its first non-empty child, and so on. 
Recall that $rank_1(T,v) \cdot K^2$ gives the position in $T$ where the children of $v$ start and each of them is represented with $K^2$ bits. We can obtain the number of children of $v$ as $NumChildren(v)=rank_1(T,v+K^2)-rank_1(T,v-1)$. Non-empty nodes store their differential encoding in $counts$ in level order, so we must be able to compute the level order of a node in constant time. Node $v$ stores $K^2$ bits that indicate which children of $v$ are non-empty. If the $i^{th}$ bit is set to 1, then the level order of that child node is $rank_1(T,v+i)$, with $i\in[0,K^2-1]$.

Given a node $v$ with absolute counting value $c_v$ and $NumChildren(v)$ children, each child $v'$ is expected to contain $\overline{count(v')}=\frac{c_v}{NumChildren(v)}$ elements. Let $p$ be the level order number of child $v'$. Then, the absolute counting value of $v'$ can be computed as $c_{v'}=counts[p]+\overline{count(v')}$. Note that $counts$ is stored using DACs, which support direct access. We use the computed value $c_{v'}$ to recursively visit the children of $v'$.

Let us consider a query example $q=[0,1]\times[0,2]$. We start at the root with $c_0 = 22$ and $result=0$. The root has three children, but only the first one, stored at position 4 in $T$, intersects $q$. Each child is expected to represent $\lfloor 22/3 \rfloor = 7$ elements, so we set $c_4=counts[rank_1(T,0+0)]+7=3+7=10$. Similarly, we recurse on the first and third child of this node. On the first branch of the recursion, we process node 16 and set $c_{16}=counts[rank_1(T,4+0)]+\lfloor 10/4 \rfloor=counts[4]+2=2$. As the submatrix corresponding with this node is contained in $q$, we add 2 to the result and stop the recursion on this branch. On the other child, we have to recurse until the leaves in order to sum the other element to the result, and obtain the final count of 3.

\subsubsection{Other supported queries}\label{subsub_other}

This data structure obviously supports all the queries that can be implemented on a $K^2$-tree, such as range reporting or emptiness queries. More interesting is that it can also support other types of queries with minor modifications. A natural generalization of range counting queries are aggregated sum queries. In this case, we consider a matrix $M[n\times n]$ where each cell can either be empty or contain a weight in the range $[0,d-1]$. We perform the same data partitioning on the conceptual binary matrix that represents the non-empty cells. In other words, we use a $K^2$-tree to represent the topology of the matrix. Then, instead of augmenting the nodes with the count of the non-empty cells in the corresponding submatrix, we store the sum of the weights contained in such submatrix. The same differential encoding used for range counting can be used to store these sums. In this case, however, the space-efficiency achieved by the data structure depends not only on the clustering of the data, but also on the distribution of the weights. The encoding achieves its best performance when the sums of the weights of all the children of a node are similar.


\section{Experiments and Results}\label{sec:experiments}

In this section we empirically evaluate the two types of queries studied in previous sections. As the datasets and evaluated solutions for both scenarios are quite different, we devote one subsection to each type of query: we first present the experiment setup (baselines and datasets), then an evaluation in terms of space usage, and finally a running time comparison.

All the data structures were implemented by ourselves and the source code is available at \url{http://lbd.udc.es/research/aggregatedRQ}. We ran all our experiments on a dedicated server with 4 Intel(R) Xeon(R)
E5520 CPU cores at 2.27GHz 8MB cache and 72GB of RAM memory. The machine
runs Ubuntu GNU/Linux version 9.10 with kernel 2.6.31-19-server (64 bits) and
gcc 4.4.1. All the data structures were implemented in C/C++ and compiled with
full optimizations.

All bitmaps that are employed use a bitmap representation that supports $rank$ and $select$ using $5\%$ of extra space. The wavelet tree employed to implement the solution of Navarro et al.\ \cite{NNR13} is a pointerless version obtained from {\sc LIBCDS} (\url{http://www.github.com/fclaude/libcds}). This wavelet tree is augmented with an RMQ data structure at each level, which requires $2.37n$ bits and solves range maximum queries in constant time.

\subsection{Ranked range queries}
\subsubsection{Experiment setup}

We use several synthetic datasets, as well as some real datasets where top-$k$ queries are of interest. 
Our synthetic datasets are square matrices where only some of the cells have a
value set. We build different matrices varying the following parameters: the
\emph{size} $s \times s$ of the matrix ($s=1024$, $2048$, $4096$, $8192$), the
number of different weights $d$ in the matrix (16, 128, 1024) and the
\emph{percentage} $p$ of cells that have a point (10, 30, 50, 70, 100\%). The
distribution of the weights in all the datasets is uniform, and the spatial
distribution of the cells with points is random. For example, the synthetic
 dataset with $(s=2048,d=128,p=30)$ has size $2048 \times 2048$, 30\% of its cells have a value and their values follow a uniform distribution in $[0,127]$. 

We also test our representation using real datasets. We extracted two
different views from a real OLAP database (\url{https://www.fupbi.com}\footnote{The dataset belongs to SkillupChile\textsuperscript{\textregistered}, which allow us to use it for our research.}) storing information about sales
achieved per store/seller each hour over several months: $salesDay$ stores the
number of sales per seller per day, and $salesHour$ the number of sales per
hour. Huge historical logs are accumulated over time, and are subject to data
mining processing for decision making. In this case, finding the places (at
various granularities) with most sales in a time period is clearly relevant. 
Table~\ref{table:real} shows a summary with basic information about the real 
datasets. For simplicity, in these datasets we ignore the cost of mapping 
between real timestamps and seller ids to rows/columns in the table, and 
assume that the queries are given in terms of rows and columns.

\begin{table}[t]
\begin{center}
\caption{Description of the real datasets used, and space (in bits per cell) required to represent them with the compared data structures. \label{table:real}
}{
\scalebox{0.8}{
\begin{tabular}{|c|r|r|r|r|r|r|}
\hline
Dataset     	&  \#Sellers		&	Time instants	&	Number
of & $K^2$-treap & $mk2tree$ & $wtrmq$ \\
				&	(rows)~			&
(columns)~~~		&	diff.\ values & (bits/cell)& (bits/cell)&
(bits/cell) \\
\hline
$SalesDay$    	&  1314				&	471
&	297 	& 2.48 & 3.75 & 9.08 \\
$SalesHour$    	&  1314				&	6028
&	158 	& 1.06 & 0.99 & 3.90 \\
\hline
\end{tabular}}
}
\end{center}
\end{table}

We compare the space requirements of the $K^2$-treap with a
solution based on wavelet trees enhanced with RMQ structures
\cite{NNR13} ($wtrmq$).
Since our matrices can contain none or multiple values per column, we
transform our datasets to store them using wavelet trees. The wavelet tree
will store a grid with as many columns as values we have in our matrix, in
column-major order. A bitmap is used to map the real columns with virtual
ones: we append a 0 per new point and a 1 when the column changes.
Hence, range queries in the $wtrmq$ require a mapping from real columns to
virtual ones (2 $select_1$ operations per query), and the virtual column of
each result must be mapped back to the actual value (a $rank_1$ operation per
result).

We also compare our proposal with a representation based on constructing
multiple $K^2$-trees, one per different value in the dataset. In this
representation ($mk2tree$), top-$k$ queries are answered by querying
consecutively the $K^2$-tree representations for the higher values. Each
$K^2$-tree representation in this proposal is enhanced with multiple
optimizations over the simple bitmap approach we use, like the compression of
the lower levels of the tree (see~\cite{ktree} for a detailed
explanation of this and other enhancements of the $K^2$-tree).

\subsubsection{Space comparison}
We start by comparing the compression achieved by the representations. As
shown in Table~\ref{table:real}, the $K^2$-treap overcomes the $wtrmq$ in
the real datasets considered by a factor over 3.5. 
Structure $mk2tree$ is competitive with
the $K^2$-treap and even obtains slightly less space in the dataset
$salesHour$, taking advantage of the relatively small number of different
values in the matrix. 

\begin{figure}[t]
\begin{center}
   \includegraphics[width=\textwidth]{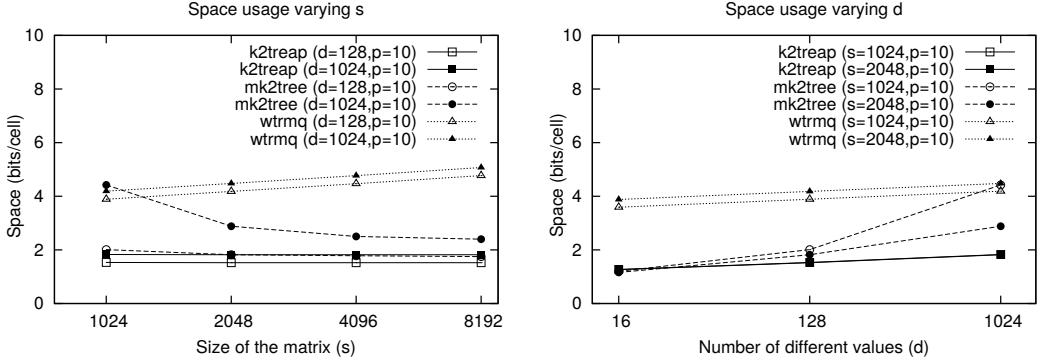}   
   \caption{Evolution of the space usage with $s$ and $d$ in the synthetic datasets, in bits/cell (in the right plot, the two lines for the $K^2$-treap coincide).}
   \label{fig:spaceSint}   
\end{center}
	
\end{figure}

The $K^2$-treap also obtains the best space
results in most of the synthetic datasets studied. Only in the datasets with 
very few different values ($d=16$) the $mk2tree$ uses less space
than the $K^2$-treap. Notice that, since the distribution of values and cells
is uniform, the synthetic datasets are close to a worst-case scenario for
the $K^2$-treap and $mk2tree$. 
Fig.~\ref{fig:spaceSint} provides a summary of the space results for some of
the synthetic datasets used. The left plot shows the evolution of compression
with the size of the matrix. The $K^2$-treap is almost unaffected by the
matrix size, as its space is around $t\log\frac{s^2}{t} = 
s^2\frac{p}{100}\log\frac{100}{p}$ bits, that is, constant per cell as $s$ 
grows. On the other hand, the $wtrmq$ uses 
$t\log s = s^2\frac{p}{100}\log s$ bits, that 
is, its space per cell grows logarithmically with $s$. Finally, the $mk2tree$ 
obtains poor results in the smaller datasets but it is more competitive on larger
ones (some enhancements in the $K^2$-tree representations behave worse in
smaller matrices). Nevertheless, notice that the improvements in the $mk2tree$
compression stall once the matrix reaches a certain size. 

The right plot of
Fig.~\ref{fig:spaceSint} shows the space results when varying the number of
different weights $d$. The $K^2$-treap and the $wtrmq$ are affected only
logarithmically by $d$. The $mk2tree$, instead, is sharply affected, since it
must build a different $K^2$-tree for each different value: if $d$ is very
small the $mk2tree$ representation obtains the best space results also in the
synthetic datasets, but for large $d$ its compression degrades significantly.

As the percentage of cells set $p$ increases, the compression in terms of bits/cell (i.e., total bits divided by $s^2$) will be worse. 
However, if we measure the compression in bits/point (i.e., total bits 
divided by $t$), then the space of the $wtrmq$ is independent of $p$ ($\log s$ 
bits), whereas the $K^2$-treap and $mk2tree$ use less space as $p$ increases
($\log \frac{100}{p}$). That is, the space usage of the $wtrmq$ increases 
linearly with $p$, while that of the $K^2$-treap and $mk2tree$ increases 
sublinearly. Over all the synthetic datasets, the $K^2$-treap uses from 1.3 to 
13 bits/cell, the $mk2tree$ from 1.2 to 19, and the
$wtrmq$ from 4 to 50 bits/cell. 

\subsubsection{Query processing}
In this section we analyze the efficiency of top-$k$ queries, comparing our
structure with the $mk2tree$ and the $wtrmq$. For each dataset, we build
multiple sets of top-$k$ queries for different values of $k$ and different spatial ranges (we ensure that the spatial ranges contain at least $k$ points). All query sets are generated for fixed $k$ and $w$ (side of the spatial window). Each query set contains 1,000 queries where the spatial window is placed at a random position within the matrix. 

\begin{figure}[t]
\begin{center}
   \includegraphics[width=\textwidth]{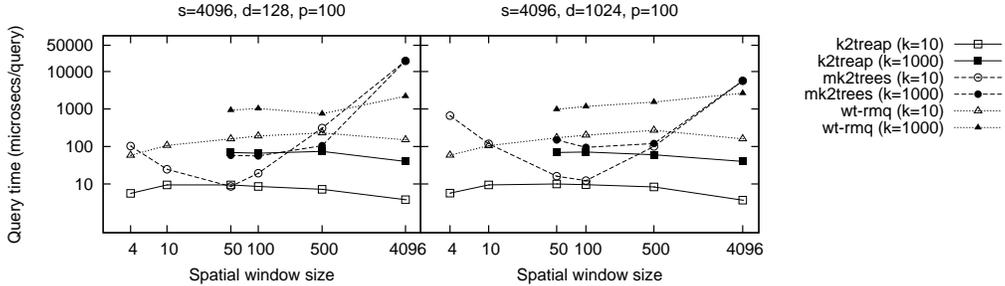}
   \caption{Times (in microseconds per query) of top-$k$ queries in synthetic datasets for $k=10$ and $k=100$ and range sizes varying from 4 to 4,096. The number of different weights $d$ in the matrix is 128 on the left graph and 1,024 on the right graph, while $s$ and $p$ remain fixed. We omit the lines connecting the points for $wtrmq$ variants, as they produce several crosses that hamper legibility.}
   \label{fig:topkSint}   
\end{center}
	
\end{figure}

Fig.~\ref{fig:topkSint} shows the time required to perform top-$k$ queries in
some of our synthetic datasets, for different values of $k$ and $w$.
The $K^2$-treap obtains better query times than the
$wtrmq$ in all the queries, and both evolve similarly with the size of the
query window. On the other hand, the $mk2tree$ representation obtains poor
results when the spatial window is small or large, but it is competitive with the
$K^2$-treap for medium-sized ranges. This is due to the procedure to query the
multiple $K^2$-tree representations: for small windows, we may need to query
many $K^2$-trees until we find $k$ results; for very large windows, the
$K^2$-treap starts returning results in the upper levels of the conceptual
tree, while the $mk2tree$ approach must reach the leaves; for some
intermediate values of the spatial window, the $K^2$-treap still needs to
perform several steps to start returning results, and the $mk2tree$
representation may find the required results in a single $K^2$-tree. Notice
that the $K^2$-treap is more efficient when no range limitations are given
(that is, when $w = s$), since it can return after exactly $K$ iterations.
Fig.~\ref{fig:topkSint} only shows the results for two of the datasets, but
similar results were obtained in all the synthetic datasets
studied.

\begin{figure}[]
	\begin{center}
	   \includegraphics[width=\textwidth]{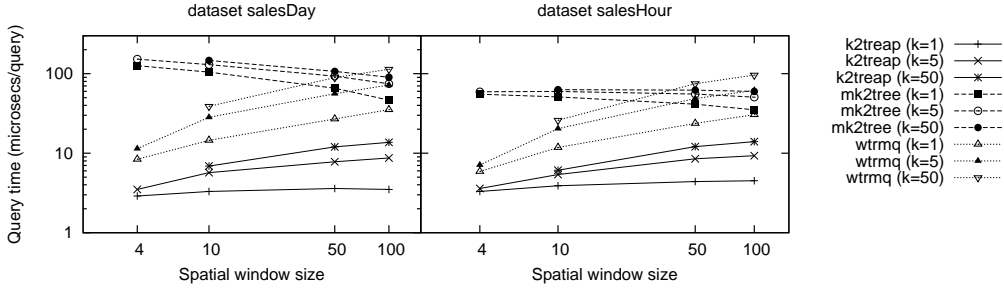}
	   \caption{Query times (in microseconds per query) of top-$k$ queries in the real datasets $SalesDay$ (left) and $salesHour$ (right) for $k=1$, $k=5$ and $k=50$, and range sizes varying from 4 to 100.}
	   \label{fig:topkRealesVentana}   
	\end{center}	
\end{figure}

Next we query our real datasets. We start with the same $w \times w$ queries 
as before, which filter a range of
rows (sellers) and columns (days/hours). Fig.~\ref{fig:topkRealesVentana} shows
the results of these range queries. As we can see, the $K^2$-treap outperforms
both the $mk2tree$ and $wtrmq$ in all cases. As in the synthetic spaces, the $mk2tree$ obtains poor query times for small ranges but it is better in larger ranges. 

\begin{figure}[h]
	\begin{center}
		\begin{minipage}[t]{0.49\textwidth}
			 \includegraphics[width=\textwidth]{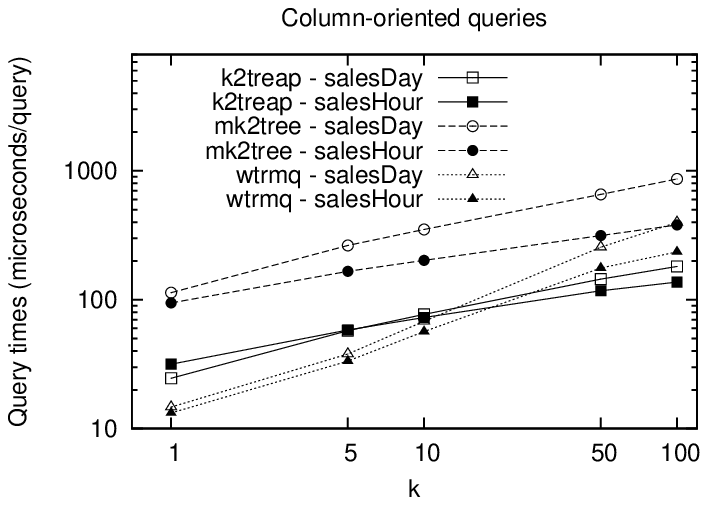}
		\end{minipage}
	  	\begin{minipage}[t]{0.49\textwidth}
			 \includegraphics[width=\textwidth]{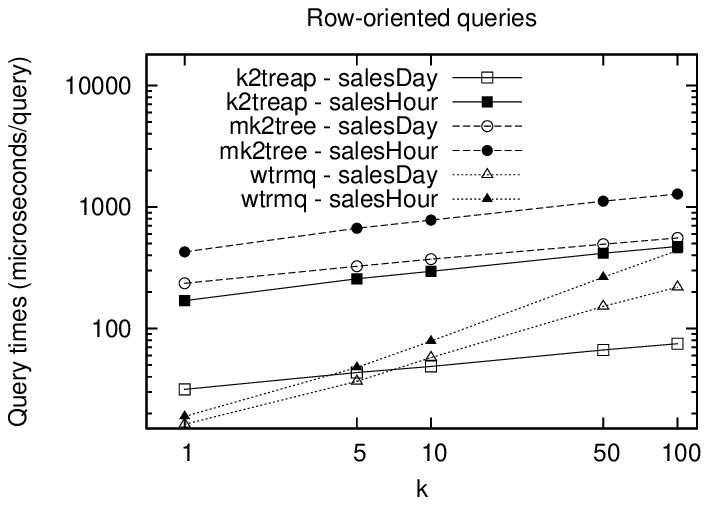}
		\end{minipage}
	   \caption{Query times (in microseconds per query) of column-oriented (left) and row-oriented (right) top-$k$ queries on the real datasets for $k$ varying from 1 to 100.}
	   \label{fig:topkRealesVentana2}   
	\end{center}	
\end{figure}

We also run two more specific sets of queries that may be of interest
in many datasets, as they restrict only the range of sellers or the time
periods, that is, only one of the dimensions
of the matrix. Row-oriented queries ask for a single row (or a small range of
rows) but do not restrict the columns, and column-oriented ask for single
columns. We build sets of 10{,}000 top-$k$ queries for random rows/columns
with different values of $k$. Fig.~\ref{fig:topkRealesVentana2} (left) shows
that in column-oriented queries the $wtrmq$ is faster than the $K^2$-treap for
small values of $k$, but our structure is still faster as $k$ grows. The reason
for this difference is that in ``square'' range queries, the $K^2$-treap only
visits a small set of submatrices that overlap the region; in
row-oriented or column-oriented queries, the $K^2$-treap is forced to check
many submatrices to find only a few results. The $mk2tree$ suffers from 
the same problem, being unable to efficiently filter the matrix, and
obtains the worst query times in all cases. 

In row-oriented queries
(Fig.~\ref{fig:topkRealesVentana2}, right) the $wtrmq$ is even more
competitive, obtaining the best results in many queries. The reason for the
differences with column-oriented queries in the
$wtrmq$ is the mapping between real and virtual columns:
column ranges are expanded to
much longer intervals in the wavelet tree, while row ranges are left
unchanged. Notice anyway that our structure is still competitive unless $k$
is very small.

\subsection{Range counting queries}
\subsubsection{Experiment setup}

In this evaluation, we use grid datasets coming from three different real domains: Geographic Information Systems (GIS), Social Networks (SN) and Web Graphs (WEB). For GIS data we use the Geonames dataset\footnote{http://www.geonames.org}, which contains the geographic coordinates (latitude and longitude) of more than 6 million populated places, and converted it into three grids with different resolutions: Geo-sparse, Geo-med, and Geo-dense. The higher the resolution, the sparser the matrix. These datasets allow for evaluating the influence of data sparsity in the different proposals. For SN we use three social networks (dblp-2011, enwiki-2013 and ljournal-2008) obtained from the Laboratory for Web Algorithmics\footnote{{http://law.di.unimi.it}}~\cite{BoVWFI,BRSLLP}. Finally, in the WEB domain we consider the grid associated with the adjacency matrix of three Web graphs (indochina-2004, uk-2002 and uk-2007-5) obtained from the same Web site. 
The clustering in these datasets is very dissimilar. In general, GIS datasets do not present many clusters, whereas data points in the WEB datasets are highly clustered. SN represents an intermediate collection in terms of clustering.

In this experiment, we compare our proposal to speed up range counting queries on a $K^2$-tree, named $rck2tree$, with the original $K^2$-tree. Recall from Section~\ref{sec:rangecounting} that we augment the $K^2$-tree with additional data in order to speed up this type of queries. Thus, in this evaluation we show the space-time trade-off offered by the $rck2tree$ structure, which is parametrized by the number of levels of the original $K^2$-tree augmented with counting information. As a baseline from the succinct data structures area, we include a representation of grids based on wavelet trees, named \emph{wtgrid} in the following discussion. This representation was described in Section \ref{subsec:wt} and an algorithm to support range counting was sketched in Section \ref{sec:relwork}. As explained in Section \ref{subsec:wt}, this representation requires a bitmap to map from a general grid to a grid with one point per column. In our experiments we store this bitmap with either plain bitmaps, RRR or sd-arrays, whichever requires less space. As for the wavelet tree itself, we use a balanced tree with just one pointer per level and the bitmaps of each level are compressed with RRR. In other words, we use a configuration of this representation that aims to reduce the space. Note, however, that we use the implementation available in {\sc Libcds} \cite{CN08}, which uses $O(alphabet\_size)$ counters to speed up queries. As we will show in the experiments, this data structure, even with the compression of the bitmaps that represent the nodes of the wavelet tree, does not take full advantage of the existence of clusters in the data points.

Table \ref{tab:space} shows the main characteristics of the datasets used: name of the dataset, size of the grid ($u$)\footnote{Note that, unlike the grids used in the previous scenario, these are square grids, and thus $u$ represents both the number of rows and columns.}, number of points it contains ($n$) and the space achieved by the baseline $wtgrid$, by the original $K^2$-tree and by different configurations of our $rck2tree$ proposal. Unlike the previous scenario, the space is measured in bits per point because these matrices are very sparse, which results in very low values of bits per cell.

\begin{table}[t]
\begin{center}
\caption{Description of the real datasets used, and space (in bits per point) required to represent them with the compared data structures.\label{tab:space}
}{
\scalebox{0.6}{
\begin{tabular}{| c | l | r | r | r | r | r | r | r |}
\hline
 Dataset  & Type & Grid (u) & Points (n) & wtgrid & $K^2$-tree &  $rck2tree^4$  &  $rck2tree^8$ & $rck2tree^{16}$\\
 			&	 &			&			&  (bits/point) &  (bits/point) &	(bits/point)		&	(bits/point)	&	(bits/point)	\\		
 \hline
Geo-dense    & GIS &     524,288 &   6,049,875 & 17.736 & 14.084	& 14.085	&	14.356	& 18.138	\\
Geo-med      & GIS &    4,194,304 &   6,080,640 & 26.588 & 26.545	&	26.564	&	29.276	&	36.875	\\
Geo-sparse   & GIS &  67,108,864 &   6,081,520 & 44.019 & 41.619	&	41.997	&	48.802	&	56.979	\\\hline
dblp-2011       & SN  &     986,324 &   6,707,236 & 19.797 & 9.839	&	9.844	&	10.935	&	13.124	\\
enwiki-2013     & SN  &   4,206,785 & 101,355,853 & 19.031 & 14.664	&	14.673	&	16.016	&	19.818	\\
ljournal-2008   & SN  &   5,363,260 & 79,023,142 & 20.126 & 13.658	&	13.673	&	15.011	&	18.076	\\\hline
indochina-2004  & WEB &   7,414,866 & 194,109,311 & 14.747 & 1.725	&	1.729	&	1.770	&	2.13	\\
uk-2002         & WEB &  18,520,486 & 298,113,762 & 16.447 & 2.779	&	2.797	&	2.888	&	3.451	\\
uk-2007-5       & WEB &  105,896,555  & 3,738,733,648  & 16.005 & 1.483	&	1.488	&	1.547	&	1.919	\\
\hline
\end{tabular}}
}
\end{center}
\end{table}

\subsubsection{Space comparison}

As expected, the representation based on wavelet trees, $wtgrid$, is not competitive in terms of space, especially when the points in the grid are clustered. Even though the nodes of the wavelet tree are compressed with RRR, this representation is not able to capture the regularities induced by the clusters. In the WEB domain, where the points are very clustered, the $wtgrid$ representation requires up to 10 times the space of the $K^2$-tree. Therefore, the latter allows for the processing in main memory of much larger datasets. In domains where the data points are not that clustered, space-savings are still significant but not as outstanding as in the previous case. 

Second, we analyze the space overhead incurred by the $rck2tree$ in comparison with the original $K^2$-tree. As mentioned above, this overhead depends on the number of levels of the $K^2$-tree that are augmented with additional range counting information. In these experiments, we show the results of three configurations in which 4, 8 and 16 levels were augmented, respectively. As Table \ref{tab:space} shows, the space overhead is almost negligible for the $rck2tree^4$ and it ranges from 25\% to 40\% for $rck2tree^{16}$. In the next section we will show the influence of this extra space in the performance of the data structure to solve range counting queries.

\begin{figure}[t]
	\begin{center}
		\begin{minipage}[t]{0.49\textwidth}
			 \includegraphics[width=\textwidth]{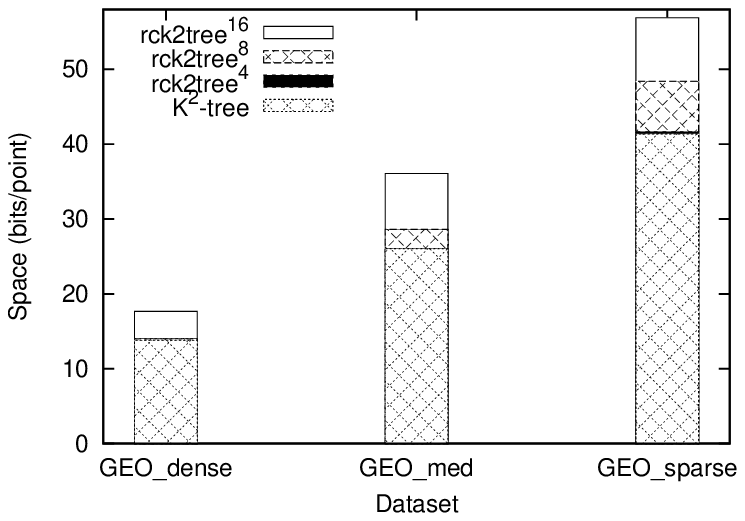}
		\end{minipage}
	  	\begin{minipage}[t]{0.49\textwidth}
			 \includegraphics[width=\textwidth]{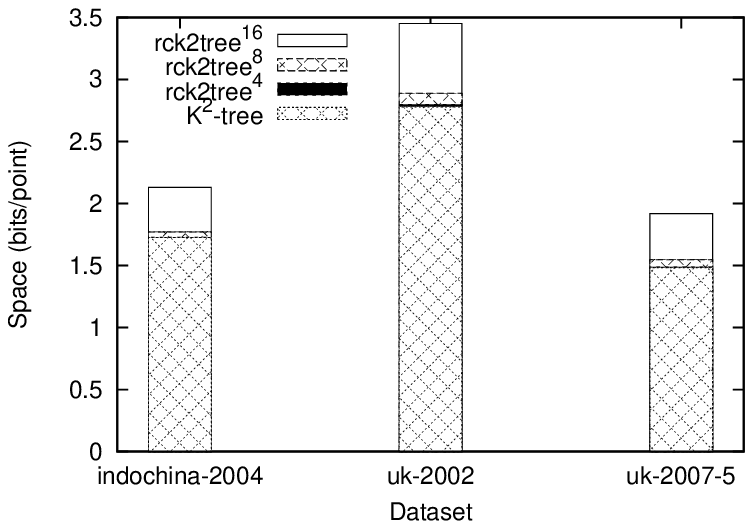}
		\end{minipage}
	   \caption{Space overhead (in bits per point) in the real datasets GIS (left) and WEB (right). Each bar, from bottom to top, represents the $K^2$-tree and the additional space used by the $rck2tree$ with 4, 8, and 16 levels, respectively. The additional space required by $rck2tree^4$ is almost negligible.}
	   \label{fig:rc_space}   
	\end{center}	
\end{figure}

It is interesting to notice that the space overhead is lower in the domains where the $K^2$-performs best. The $K^2$-tree performs better in sparse domains where data points are clustered \cite{ktree}, for example, in the Web graph. In the largest WEB dataset, uk-2007-5, the $K^2$-tree requires about 1.5 bits per point, and we can augment the whole tree with range counting information using less than 0.5 extra bits per point (this is an overhead of less than 30\%). Sparse and clustered matrices result in less values being stored in the augmented data structure (as in the original $K^2$-tree, we do not need to store information for submatrices full of zeros). In Fig. \ref{fig:rc_space} we show the space overhead in two of our dataset collections, GIS and WEB.

Note also that, unlike the $K^2$-tree, the space overhead does not increase drastically in sparse non-clustered datasets (e.g., Geo-sparse). This is because isolated points waste $K^2$ bits per level in the original $K^2$-tree, and this is much more than the overhead incurred by the range counting fields, where they use approximately two bits per level. The reason is that these fields represent the difference between the expected number of elements in the submatrix, 1, and the actual number, which is also 1.\footnote{Recall that these data are represented using DACs, which require at least two bits.}. For example, in Geo-sparse the $K^2$-tree uses more than 40 bits per point, which is much more than the (roughly) 2 bits per point in the sparse and clustered WEB datasets. However, the space overhead of the $rck2tree^{16}$ (with respect to the $K^2$-tree) is about 40\% in Geo-sparse and 30\% in uk-2007-5 (i.e., a difference of 10 percentage points). In the configurations that store range counting values only for some levels of the $K^2$-tree, the difference is even smaller. This is expected because there are fewer isolated points in the higher levels.

\subsubsection{Query processing}

In this section we analyze the efficiency of range counting queries, comparing our augmented $K^2$-tree with the original data structure and with the $wtgrid$. For each dataset, we build multiple sets of queries with different query selectivities (i.e., size of spatial ranges). All the query sets were generated for fixed query selectivity. A query selectivity of $X\%$ means that each query in the set covers $X\%$ of the cells in the dataset. Each query set contains 1,000 queries where the spatial window is placed at a random position in the matrix. 

\begin{figure}[]
	\begin{center}
		\begin{minipage}[t]{0.49\textwidth}
			 \includegraphics[width=\textwidth]{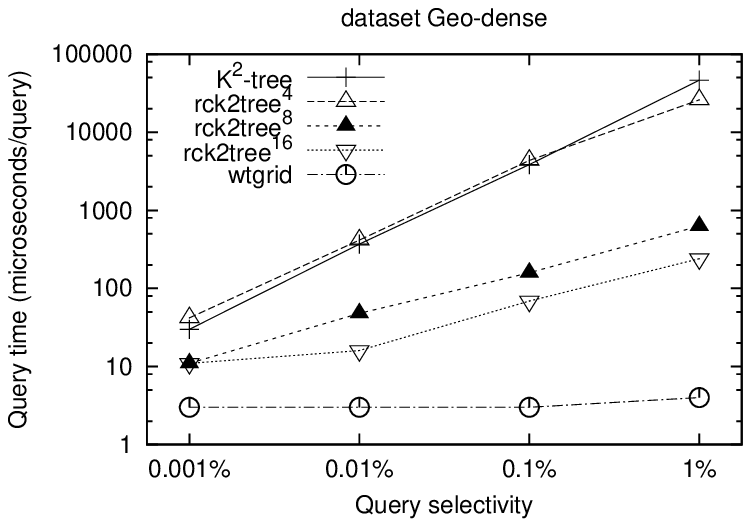}
		\end{minipage}
	  	\begin{minipage}[t]{0.49\textwidth}
			 \includegraphics[width=\textwidth]{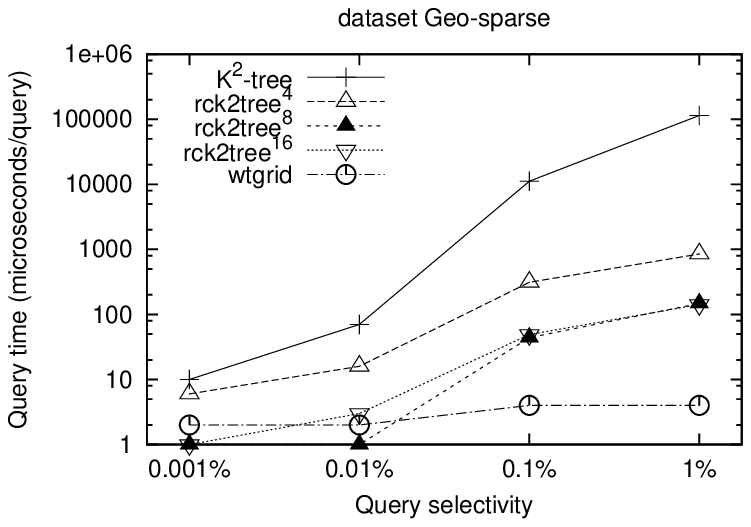}
		\end{minipage}
		\begin{minipage}[t]{0.49\textwidth}
			 \includegraphics[width=\textwidth]{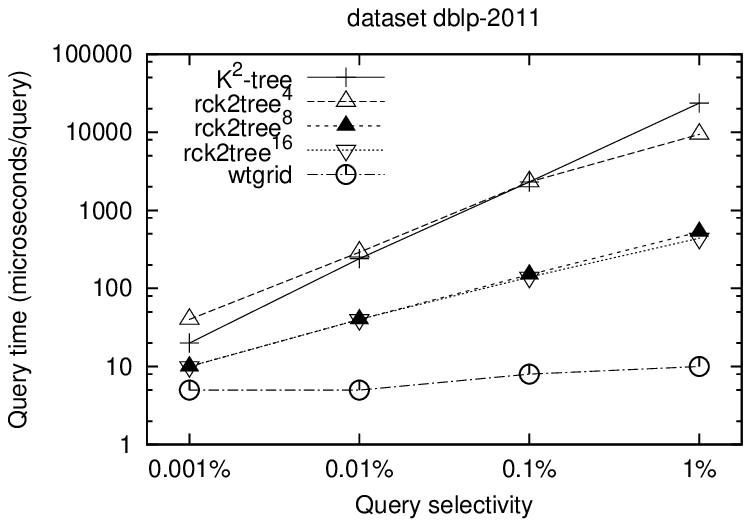}
		\end{minipage}
	  	\begin{minipage}[t]{0.49\textwidth}
			 \includegraphics[width=\textwidth]{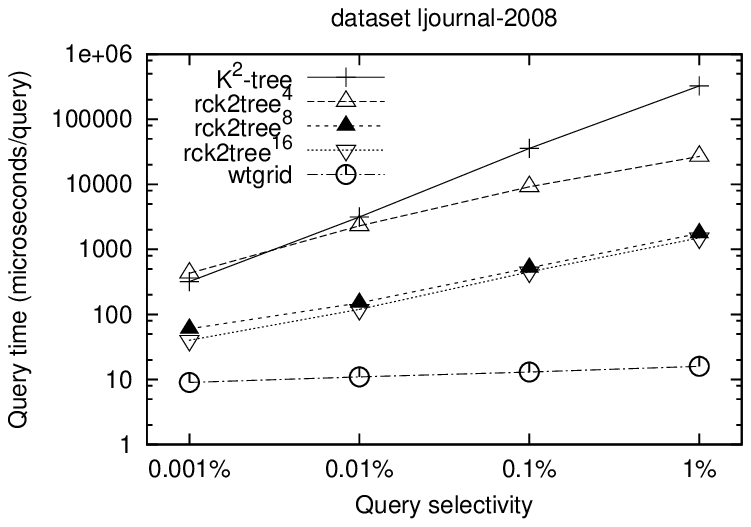}
		\end{minipage}
		\begin{minipage}[t]{0.49\textwidth}
			 \includegraphics[width=\textwidth]{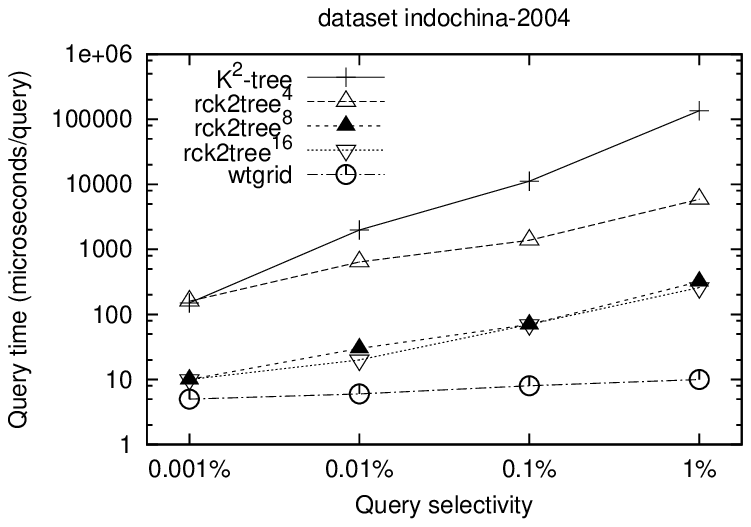}
		\end{minipage}
	  	\begin{minipage}[t]{0.49\textwidth}
			 \includegraphics[width=\textwidth]{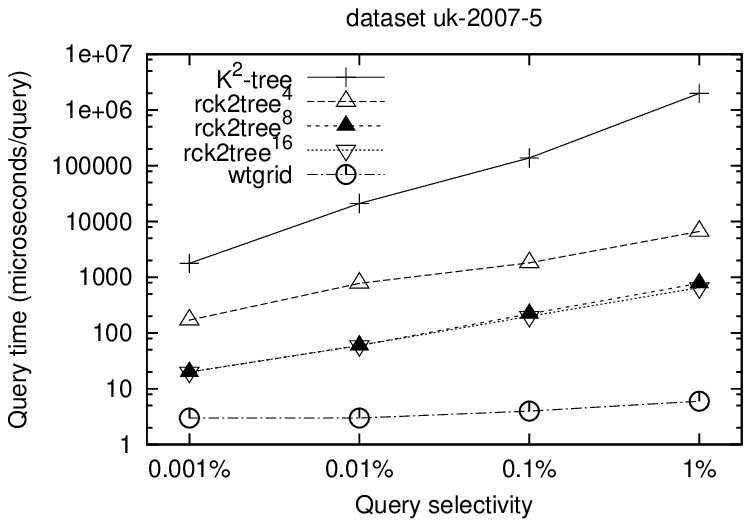}
		\end{minipage}
	   \caption{Query times (in microseconds per query) of range counting queries in two examples of each type of real dataset: GIS (top), SN (middle) and WEB (bottom) for range queries sizes varying from $0.001\%$ to $1\%$.}
	   \label{fig:rc_time}   
	\end{center}	
\end{figure}

As in the space comparison, we show the results of three different configurations of our augmented data structure, in which 4, 8, and 16 levels are augmented with range counting data. Fig. \ref{fig:rc_time} shows the time required to perform range counting queries in some of our real datasets, for different values of query selectivity. For each domain (GIS, SN and WEB), we only show the results of the two most different datasets, as the others do not alter our conclusions.

Our augmented data structure consistently outperforms the original $K^2$-tree for all domains and query selectivities. The only exception is the $rck2tree^4$ in the two SN datasets and Geo-dense for very selective queries (i.e., with the smallest areas). The influence of the query selectivity in the results is evident. The larger the query, the higher the impact of the additional range counting data in the performance of the data structure. Larger queries are expected to stop the recursion of the range counting algorithm in higher levels of the $K^2$-tree because those queries are more likely to contain the whole area covered by nodes of the $K^2$-tree. Recall that a node of the $K^2$-tree at level $i$ represents $u^2/(K^2)^i$ cells. In our experiments we use a configuration of the $K^2$-tree in which the first six levels use a value of $K_1=4$ (thus partitioning the space into $K_1^2=16$ regions) and the remaining levels use a value of $K_2=2$. Therefore, for the $rck2tree^4$ to improve the performance of the range counting queries, those queries must contain at least $u^2/(4^2)^4= u^2/2^{16}$ cells. In the $rck2tree^8$ and $rck2tree^{16}$, which contain range counting data for more levels of the $K^2$-tree, this value is much lower, and thus the recursion can be stopped early even for small queries. For larger queries, the performance improvement reaches several orders of magnitude in some datasets (note the logarithmic scale).

In most datasets, the performance of $rck2tree^8$ and $rck2tree^{16}$ is very similar, therefore the former is preferable as it requires less extra space. Hence, in general, we can recommend the use of the $rck2tree^8$ configuration to speed up range counting queries. If the queries are not very selective, and there are memory constraints, the $rck2tree^4$ can be an alternative as it requires almost the same space of the original $K^2$-tree. 

\begin{figure}[t]
	\begin{center}
		\begin{minipage}[t]{0.49\textwidth}
			 \includegraphics[width=\textwidth]{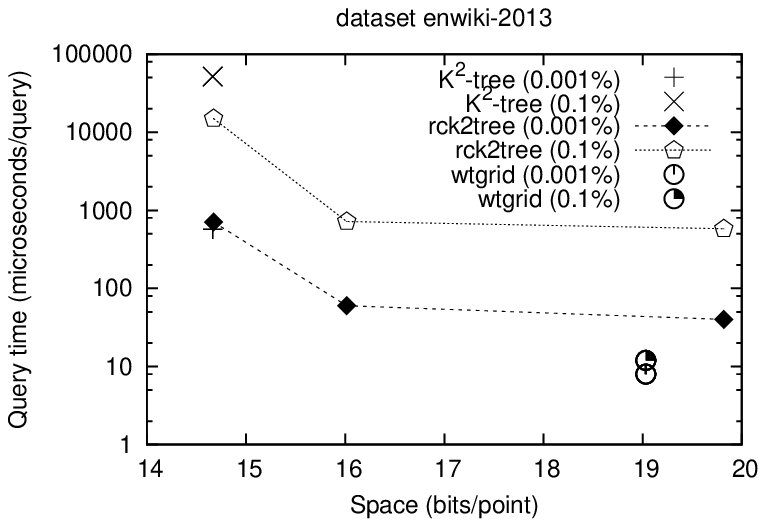}
		\end{minipage}
	  	\begin{minipage}[t]{0.49\textwidth}
			 \includegraphics[width=\textwidth]{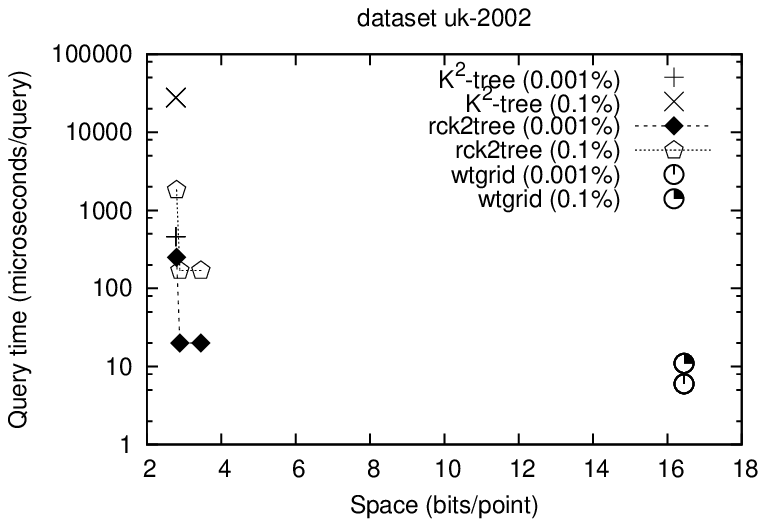}
		\end{minipage}
	   \caption{Space-time trade-offs offered by the compared range counting variants on examples of SN (left) and WEB (right) datasets.}
	   \label{fig:rc_tradeoff}   
	\end{center}	
\end{figure}

The $wtgrid$ is consistently not only faster than the data structures based on the $K^2$-tree, but also less sensitive to the size of the query. However, as we showed above, it also requires significantly more space. To reinforce these conclusions, Fig. \ref{fig:rc_tradeoff} shows the space-time trade-offs of the different configurations. We selected two representative datasets that were not used in the previous experiment, enwiki-2013 (SN) and uk-2002 (WEB), and two query selectivities for each of them, $0.001\%$ and $0.1\%$. Each point in the lines named $rck2tree$ represents a different configuration with 4, 8, and 16 levels augmented with range counting information, respectively from left to right.

These graphs show that the $rck2tree^8$ configuration offers the most interesting trade-off, as it requires just a bit more space than the original $K^2$-tree and it speeds up range counting queries significantly. This effect is more evident for larger queries, but even for very selective queries the improvement is significant. The $wtgrid$ competes in a completely different area of the trade-off, being the fastest data structure, but also the one that requires the most space.

\section{Conclusions}\label{sec:conclusions}

We have introduced a technique to solve aggregated 2D range queries on grids, which requires little space when the data points in the grid are clustered. We use a $K^2$-tree to represent the topology of the data (i.e., the positions of the data points) and augment each node of the tree with additional aggregated information that depends on the operation to be supported. The aggregated information in each node is differentially encoded with respect to its parent in order to reduce the space overhead. To illustrate the applicability of this technique, we adapted it to support two important types of aggregated range queries: ranked and counting range queries.

In the case of ranked queries, we named the resulting data structure $K^2$-treap. This data structure performs top-$k$ range queries up to 10 times faster than current state-of-the-art
solutions and requires as little as 30\% of their space, both in synthetic and real OLAP datasets. This holds even on uniform distributions, which is the worst scenario for $K^2$-treaps.

For range counting queries, our experimental evaluation shows that with a small space overhead (below 30\%) on top of the $K^2$-tree, our data structure answers queries several orders of magnitude faster than the original $K^2$-tree, especially when the query ranges are large. These results are consistent in the different databases tested, which included domains with different levels of clustering in the data points. For example, in Web graphs the data points are very clustered, which is not the case in GIS applications. The comparison with a wavelet tree-based solution shows that, although the wavelet tree is faster, our proposal requires less space (up to 10 times less when the points are clustered). Thus, we provide a new alternative in the space-time trade-off which allows for the processing of much larger datasets.

Although we have presented the two types of queries separately, this does not mean that an independent data structure would be required for each type of aggregated query. The topology of the data can be represented by a unique $K^2$-tree and each type of aggregated query just adds additional aggregated and differentially encoded information. However, some specific optimizations on the $K^2$-tree, such as the one presented for ranked range queries, may not be possible for all types of queries.

The technique can be generalized to represent grids in higher dimensions, which is essential in some domains such as OLAP databases \cite{Sarawagi97}, by 
replacing our underlying $K^2$-tree with its generalization to $d$
dimensions, the $K^d$-tree \cite{ABNP13} (not to be confused with $kd$-trees 
\cite{Ben75}). The algorithms stay identical, but an empirical evaluation is
left for future work. In the worst case, a grid of $t$ points on $[n]^d$ will 
require $O(t \log \frac{n^d}{t})$ bits, which is of the same order of the data,
and much less space would be used on clustered data. Instead, an extension of the
wavelet tree will require $O(n\log^d n)$ bits, which quickly becomes
impractical. Indeed, any structure able to report the points in a range in
polylogarithmic time requires $\Omega(n(\log n/\log\log n)^{d-1})$ words of
space \cite{Cha90}, and with polylogarithmic space one needs time at least
$\Omega(\log n(\log n/\log\log n)^{\lfloor d/2\rfloor-2})$ \cite{AAL12}.
As with top-$k$ queries one can report all the points in a range,
there is no hope to obtain good worst-case time and space bounds 
in high dimensions, and thus heuristics like $K^d$-treaps are the only
practical approaches ($kd$-trees do offer linear space, but their time
guarantee is rather loose, $O(n^{1-1/d})$ for $n$ points on $[n]^d$).

\bibliographystyle{elsarticle-num} 
\bibliography{paper}


\end{document}